\documentclass[letterpaper, 10 pt, conference]{ieeeconf}

\IEEEoverridecommandlockouts                              

\usepackage{graphicx} 
\usepackage{amsmath} 
\usepackage{amssymb}  
\usepackage{pifont}
\usepackage{bm}
\usepackage{cite}
\usepackage{xcolor}

\usepackage{hyperref}

\newcommand{\doi}[1]{{doi:~\href{https://doi.org/#1}{\nolinkurl{#1}}}\rmFullStop}

\newcommand*{\rmFullStop}{\rmifnextchar{.}{}{}}

\makeatletter
\newcommand{\rmifnextchar}[3]{%
  \begingroup
  \ltx@LocToksA{\endgroup#2}%
  \ltx@LocToksB{\endgroup#3}%
  \ltx@ifnextchar{#1}{%
    \def\next{\the\ltx@LocToksA}%
    \afterassignment\next
    \let\scratch= %
  }{%
    \the\ltx@LocToksB
  }%
}
\makeatother

\usepackage{subcaption}
\usepackage{epstopdf}
\usepackage{mathtools}
\usepackage{tikz}
\usetikzlibrary{patterns,positioning,shapes,calc}
\usepackage{cellspace}
\usepackage{environ}
\makeatletter
\newsavebox{\measure@tikzpicture}
\NewEnviron{scaletikzpicturetowidth}[1]{%
  \def\tikz@width{#1}%
  \begin{lrbox}{\measure@tikzpicture}%
  \BODY
  \end{lrbox}%
  \pgfmathparse{#1/\wd\measure@tikzpicture}%
  \BODY
}
\makeatother
\usepackage{array}
\newcolumntype{x}[1]{>{\centering\arraybackslash\hspace{0pt}}p{#1}}
\definecolor{mygray}{rgb}{0.7,0.7,0.7}
\definecolor{mygrouping}{HTML}{000000}
\definecolor{myred}{HTML}{c62828}
\definecolor{myquad}{HTML}{388e3c}
\definecolor{mynonlinear}{HTML}{6a1b9a}

\usepackage{blindtext}
\usepackage{ifthen}
\usepackage{hyperref}
\hypersetup{colorlinks=true, allcolors = blue,
pdftitle={Site-dependent Solutions of Wave Energy Converter Farms with Surrogate Models, Control Co-design, and Layout Optimization},
pdfauthor={Saeed Azad; Daniel R. Herber, Suraj Khanal, Gaofeng Jia},
}
\usepackage{soul}

\newcommand{\parm}{\mathord{\color{black!33}\bullet}}%
\newcommand{\RNum}[1]{\uppercase\expandafter{\romannumeral #1\relax}}
\newcolumntype{s}{>{\columncolor[HTML]{F5F5F5}} c}
\newcommand{\MyBold}[1]{\mathbf{#1}}
\newcommand{\Rwec}{R_{\text{wec}}}            
\newcommand{\ARwec}{AR_{\text{wec}}}            
\newcommand{\Dwec}{D_{\text{wec}}}            
\newcommand{\Rwecmax}{\bar{R}_{\text{wec}}}   
\newcommand{\Dwecmax}{\bar{D}_{\text{wec}}}   
\newcommand{\ARwecmax}{\bar{AR}_{\text{wec}}}   
\newcommand{\Rwecmin}{\underaccent{\bar}{R}_{\text{wec}}}   
\newcommand{\Dwecmin}{\underaccent{\bar}{D}_{\text{wec}}}   
\newcommand{\ARwecmin}{\underaccent{\bar}{AR}_{\text{wec}}}   
\newcommand{\Angpq}{\theta_{pq}}              
\newcommand{\Dispq}{l_{pq}}                   
\newcommand{\Nwec}{n_{\text{wec}}}     
\newcommand{\Mass}{\MyBold{M}}       
\newcommand{\Force}{\MyBold{F}}      
\newcommand{\AddedMass}{\MyBold{A}}     
\newcommand{\DampingCoeff}{\MyBold{B}}  
\newcommand{\OMF}{(\omega)}             
\newcommand{\AL}{\MyBold{w}}           
\newcommand{\KPTO}{\MyBold{K}_{\text{pto}}}           
\newcommand{\BPTO}{\MyBold{B}_{\text{pto}}}           
\newcommand{\KPTOmin}{\underaccent{\bar}{\MyBold{k}}_{\text{pto}}}           
\newcommand{\BPTOmin}{\underaccent{\bar}{\MyBold{B}}_{\text{pto}}}           
\newcommand{\KPTOmax}{\bar{\MyBold{k}}_{\text{pto}}}           
\newcommand{\BPTOmax}{\bar{\MyBold{B}}_{\text{pto}}}           

\usepackage{accents}
\usepackage{units}
\newcolumntype{s}{>{\columncolor[HTML]{F5F5F5}} c}
\makeatletter
\newcommand*{\rom}[1]{\expandafter\@slowromancap\romannumeral #1@}
\makeatother

\usepackage{multirow}
\usepackage{threeparttable}
\usepackage{booktabs} 
\usepackage{tablefootnote} 

\usepackage[normalem]{ulem}

\usepackage{booktabs}
\usepackage{tabularx}
\newcolumntype{Y}{>{\centering\arraybackslash}X}

\title{\LARGE \bf
Site-dependent Solutions of Wave Energy Converter Farms with Surrogate Models, Control Co-design, and Layout Optimization
}

\author{Saeed Azad$^{1\dagger}$,~Daniel R. Herber$^{1}$,~Suraj Khanal$^{1}$,~Gaofeng Jia$^{1}$
\thanks{$^{1}$~Colorado State University, Fort Collins, CO 80523}%
\thanks{$^{\dagger}$~Corresp. Author, \href{mailto:saeed.azad@colostate.edu}{saeed.azad@colostate.edu}}%
}%

\begin{document}

\newcounter{clearpageflag}
\setcounter{clearpageflag}{0}

\maketitle
\thispagestyle{empty}
\pagestyle{empty}

\begin{abstract}%
Design of wave energy converter farms entails multiple domains that are coupled, and thus, their concurrent representation and consideration in early-stage design optimization has the potential to offer new insights and promising solutions with improved performance.
Concurrent optimization of physical attributes (e.g.,~plant) and the control system design is often known as control co-design or CCD.
To further improve performance, the layout of the farm must be carefully optimized in order to ensure that constructive effects from hydrodynamic interactions are leveraged, while destructive effects are avoided. 
The variations in the joint probability distribution of waves, stemming from distinct site locations, affect the farm's performance and can potentially influence decisions regarding optimal plant selection, control strategies, and layout configurations. 
Therefore, this paper undertakes a concurrent exploration of control co-design and layout optimization for a farm comprising five devices, modeled as heaving cylinders in the frequency domain, situated across four distinct site locations: Alaskan Coasts, East Coast, Pacific Islands, and West Coast.
The challenge of efficiently and accurately estimating hydrodynamic coefficients within the optimization loop was mitigated through the application of surrogate modeling and many-body expansion principles.
Results indicate the optimized solutions exhibit variations in plant, control, and layout for each candidate site, signifying the importance of system-level design with environmental considerations from the early stages of the design process.    

\end{abstract}

\section{Introduction}\label{sec:introduction}

Due to the promise of wave energy as a viable renewable resource, the design of wave energy converter (WEC) devices has gained a lot of significance in recent years.
While technological diversity has made it challenging for scientists to form a consensus on the particular type of technology to focus on, there is a shared agreement about the necessity for increased research and investment to improve the technology readiness level of these devices.

The sizing (e.g.,~plant) and power take-off (PTO) control of WEC devices, which are known to be coupled, are expected to have significant impacts on the performance of WEC devices. 
However, only a few studies in WEC literature account for this coupling \cite{herber2013wave, pena2022control}.  
To meet the energy demands of today's and future society, make intelligent use of available resources, and reduce operational and maintenance costs, WECs must be deployed in a farm setting.
Within the WEC farm, devices interact with the incident wave and one another.
This interaction, which in linear potential flow theory is contributed to the scattering and radiation forces, is responsible for creating a constructive or destructive effect, indicating that maximizing power generation at the farm level requires careful deployment of WECs within the farm through layout optimization. 
However, layout optimization alone, without considering the impact of plant and PTO control, naively overlooks the coupling between these dominant domains and may result in suboptimal solutions  \cite{ringwood2023empowering}.  
Finally, the performance of a farm is directly influenced by its location (i.e., site location), as it determines crucial environmental factors like water depth, wave direction, significant wave height, wave period, and their joint probability distribution. 
This coupling motivates the consideration of environmental inputs within early-stage design efforts. 
The importance of this coupling between plant, control, and the environment has been highlighted in recent work on an energy-harvesting kite system~\cite{vermillion2023final}.

Therefore, the coupling between plant, control, layout, and environment must be acknowledged and investigated from the early stages of system development and design optimization to reveal superior solutions that are not accessible through conventional design practices. 
Therefore, this paper leverages the coupling that exists between plant, control, layout, and site selection domains in a concurrent optimization problem.
These investigations are enabled by estimating the hydrodynamic interaction effect through surrogate models that were developed in a previous study \cite{azad2023concurrent}.%

{Specifically, we investigate three case studies based on certain assumptions and domain coupling for a WEC farm consisting of five geometrically identical devices (i.e.,~a set of homogeneous individual plants).} 
The WECs are modeled as half-submerged cylindrical, heaving buoys in the frequency domain with a simplified linear PTO system, subject to probabilistic irregular ocean waves.  
Four representative site locations in Alaska Coasts, East Coast, Pacific Islands, and West Coast were investigated within this system-level framework.
Variations in significant wave height and wave period with their joint probability distribution are accounted for by using information from historical data.

At its core, this integrated approach requires the efficient calculation of hydrodynamic coefficients, typically calculated through a boundary element method (BEM) or the numerical multiple scattering (MS) approach. 
However, direct calculation of these quantities within an optimization loop is often computationally prohibitive.
Similar to the previous work in ~\cite{azad2023concurrent}, we address this issue by utilizing artificial neural networks (ANNs) within the framework of many-body expansion (MBE).
Using training data developed through MS, we trained and validated a series of surrogate models to efficiently estimate the hydrodynamic interaction effect, up to second degree, using MBE principles.
Further details about the development of these models, and the principles of MBE can be found in~\cite{azad2023concurrent}.
Compared to the previous study, the surrogate models in this work have been further improved by using a selective sampling strategy, which adaptively improves the accuracy of the surrogate model based on the uncertainty in the predictions established using the concept of query by committee (QBC) \cite{Seung1992}.

The remainder of this paper is organized as follows.
Sec.~\ref{sec:section2} briefly describes backgrounds on the dynamics and control of WEC farms, power calculations, array considerations, estimation of hydrodynamic coefficients, and wave climate.
Sec.~\ref{sec:section3} discusses the validation of surrogate models for the problem across the input domain.
The problem formulation, along with the results of the investigation, are discussed in Sec.~\ref{sec:section4}. Finally, Sec.~\ref{sec:conclusion} presents conclusions and limitations of the current study, as well as potential future work.

\ifthenelse{\value{clearpageflag}>0}{\clearpage}{}
\section{Background}\label{sec:section2}

\subsection{Dynamics and Control of WECs}\label{subsec:Dynamicand Control}

Under certain assumptions \cite{ning2022modelling}, linear potential flow theory holds and for a regular wave with frequency $\omega$ and unit amplitude as an input, the equation of motion for $n_{wec}$ buoys in the frequency domain can be described by:
\begin{align}
\label{eqn:EquationofMotion1}
     -{\omega}^{2} \Mass \hat{\bm{\xi}} = \hat{\Force}_{\text{FK}} + \hat{\Force}_{\text{s}} + \hat{\Force}_{\text{r}} +  \hat{\Force}_{\text{hs}} + \hat{\Force}_{\text{pto}}
\end{align}
\noindent
where $\hat{\parm}$ is the complex amplitude of $\parm$, $\hat{\bm{\xi}}(\cdot) \in \mathbb{R}^{\Nwec \times 1}$ is the displacement vector, and $\Mass \in \mathbb{R}^{\Nwec \times \Nwec}$ is the diagonal mass matrix.
The forces involved in the equation of motion are described by $\hat{\Force}_{}(\cdot)\in \mathbb{R}^{\Nwec \times 1}$, and include Froude-Krylov ($\text{FK}$), scattering ($\text{s}$), radiation ($\text{r}$), hydrostatic ($\text{hs}$), and PTO ($\text{pto}$) force.  
The excitation force is the sum of Froude-Krylov and scattering forces: $\hat{\Force}_{\text{e}}\OMF = \hat{\Force}_{\text{FK}}\OMF + \hat{\Force}_{\text{s}}\OMF$.
The radiation force, characterized by the added mass coefficient $\AddedMass(\cdot)$, and
hydrodynamic damping coefficient $\DampingCoeff(\cdot)$ matrices, is described as: $\hat{\Force}_{\text{r}}\OMF = {\omega}^{2}\AddedMass\OMF\hat{\xi}\OMF -i{\omega} \DampingCoeff\OMF\hat{\bm{\xi}}\OMF$.
The inertial increase due to water displacement as a result of the WEC motion is captured through the reactive effect in the added mass matrix, while the dissipated energy transmitted from WEC motion to the water (propagating away from the body) is captured through the damping coefficient \cite{folley2016numerical}.
$\AddedMass(\cdot)$, $\DampingCoeff(\cdot)$, and $\hat{\Force}_{\text{e}}\OMF$ constitute the set of hydrodynamic coefficients that depend on the geometry of the WEC device (i.e.,~plant), the layout, the number of WECs in the farm, and the water depth.  
The hydrostatic force is determined by the equilibrium between buoyancy and gravity: $\hat{\Force}_{\text{hs}}\OMF = - G \hat{\bm{\xi}}\OMF$
where $G = \rho gS$ is the hydrostatic coefficient.
In this equation, $S$ is the cross-sectional area at the undisturbed sea level, which directly depends on the radius of the WEC device: $\pi \Rwec^2$.
In this article, $\Rwec$, along with $\ARwec$, which is the slenderness ratio (i.e.,~radius to draft ratio), constitute the vector of plant optimization variables. 

The PTO system converts mechanical motion into electricity and is a reactive controller that enables a bidirectional power flow between the PTO spring and the buoy \cite{ning2022modelling}.
In its linear form, the PTO force is composed of two parameters,
$\KPTO \in \mathbb{R}^{\Nwec \times \Nwec}$, which is a diagonal matrix of PTO stiffness, and $\BPTO \in \mathbb{R}^{\Nwec \times \Nwec}$ , which is a diagonal matrix of PTO damping: $  \hat{\Force}_{\text{pto}}({\omega}) = - \KPTO \hat{\bm{\xi}}{(\omega)} -i{\omega}\BPTO\hat{\bm{\xi}}{(\omega)} $.

In this study, $\KPTO$ and $\BPTO$ constitute the set of control optimization variables for each individual WEC.
With all the forces defined, the complex amplitude of the motions of all WEC devices can be captured through a transfer function matrix ${\hat{\bm{\xi}}(\omega)} = {\mathbf{H}(\omega)\hat{\Force}_{\text{e}}({\omega})}$:
\begin{equation}
\footnotesize
\mathbf{H}(\omega) = \left [ [{\omega}^2(\Mass+\AddedMass({\omega})) + G + \KPTO] + i{\omega}(\DampingCoeff({\omega}) + \BPTO)  \right ]^{-1} \label{eq:TF_denom}
\end{equation}
For a regular wave with $\omega$, the time-average absorbed mechanical power is calculated as:
\begin{align}
    {\mathbf{p}_{m}(\omega) = \frac{1}{2}{\omega}^2 \hat{\bm{\xi}}^{T}\BPTO \hat{\bm{\xi}} }  \label{eq:pto_power1}
\end{align}
\noindent
At first, we need to integrate the product of wave spectrum $S_{\text{JS}}(\cdot)$ (represented through a JONSWAP spectrum) with the time-averaged power over all frequencies \cite{neshat2022layout, borgarino2012impact}:
\begin{align}
    {\mathbf{p}_{i}(H_{s}, T_{p}) = \int_{0}^{\infty}2S_{\text{JS}}(\omega |H_{s},T_{p})\mathbf{p}_{m}(\omega)d\omega } \label{eq:pto_power2}
\end{align}
where $S_{\text{JS}}(\omega |H_{s},T_{p})$ is represented as a function of frequency $\omega$, significant wave height $H_{s}$, and peak period $T_{p}$;
$\mathbf{p}_{i}(H_{s}, T_{p})$ is the mechanical power matrix.
Considering all sea states (which are discretized by $n_{gq}$ Gauss quadrature points), this equation is estimated as \cite{mercade2017layout}:
\begin{align}
    \mathbf{p}_{i}(H_{s}, T_{p}) &= {\sum_{k = 0}^{n_{w}} 2 \Delta\omega_{k} S_{JS}(\omega_k |H_{s},T_{p})\mathbf{p}_{m}(\omega)}  \label{eq:pto_power3} 
\end{align}
where $n_{w}$ is the number of frequencies in the discretized form.
Considering the lifetime of $n_{yr}$ for the farm, and the associated probability matrices of the site, which are obtained through a Gauss quadrature approach along with a Kernel smoothing function for associated historical data from the site~\cite{storlazzi2015future}, the average power can be calculated as \cite{neary2014methodology}:
\begin{align}
    {p_{a} = \eta_{\text{pcc}}\eta_{\text{oa}}\eta_{\text{t}}\sum_{y = 1}^{n_{yr}}\mathbf{p}_{i}(H_{s}, T_{p}) \mathbf{p}_{r}^{L}(H_{s},T_{p}| y)}
\end{align}
where $p_{a}$ is the average power.
The joint probability distribution of the wave climate for site $L$ in the $y$th year of the study is described by $\mathbf{p}_{r}^{L}(H_{s},T_{p}| y)$.
Further details regarding the calculation of the probability matrix is provided in Sec.~\ref{subsec:Wave}.
In this equation, $\eta_{\text{pcc}}$ is the efficiency of the power conversion chain, $\eta_{\text{oa}}$ is the operational availability, and $\eta_{\text{t}}$ is the transmission efficiency.
A simple objective function that considers the average power per unit volume of the wave-absorbing body \cite{falnes2020ocean} is used: $p_{v} = p_{a}/(\Nwec\pi \Rwec^2 \Dwec)$,
where $p_{v}$ is the average power per unit volume of the device, and $\Dwec$ is the draft length of the heaving cylinder.

\vspace{-6pt}
\subsection{Array Considerations}\label{subsec:Array}
For an array of $\Nwec$ WEC devices, a $2-$by-$\Nwec$ dimensional layout matrix $\AL = [\mathbf{w}_{1}, \mathbf{w}_{2}, \cdots, \mathbf{w}_{\Nwec}]$ is used to fully characterize the location of each device.
Each element of $\AL$ is composed of the center coordinates of the associated body in the Cartesian coordinate system. 
The relative distance and angle between any pair of WEC devices, namely $p$th and $q$th bodies, is described as $\Dispq$ and $\Angpq$, respectively.

\vspace{-6pt}
\subsection{Estimation of Hydrodynamic Coefficients}
\label{subsec:HydrodynamicEstimation}
In this study, we use ANNs within the framework of MBE from \cite{azad2023concurrent}.
The hydrodynamic interaction effect is truncated at second order; therefore, only interaction effects from all single devices and the pairs are considered in this study.

For single devices, 4 ANNs associated with added mass $A$, damping coefficient $B$, and the real and imaginary parts of the excitation force, $\mathbb{R}(\Force_{\text{e}})$ and $\mathbb{I}(\Force_{\text{e}})$, must be trained.
For pairs, in addition to these factors, the interaction effect for added mass and damping coefficient must be characterized through the surrogate model, resulting in a total of $6$ ANNs.
The surrogate models were trained using a selective sampling strategy based on QBC \cite{Seung1992}, in which a committee of learners (models) are trained on the current training data set.
The training data set is then augmented by new samples/data, which correspond to inputs where the committee has the highest level of disagreement, meaning there is a lot of uncertainty in the predictions from the trained ANNs.
The committee of ANNs is established by random initialization of weights and bias terms when training the ANNs, in addition to training each model on different sub-samples of the current data set.
The resulting surrogate models were then used within the framework of MBE to predict the hydrodynamic coefficients efficiently.
Further details about surrogate modeling and MBE can be found in~\cite{azad2023concurrent, zhang2020surrogate}.

\subsection{Wave Climate and Modeling with Site Considerations}
\label{subsec:Wave}
\begin{figure}
    \centering
    \includegraphics[width=\columnwidth]{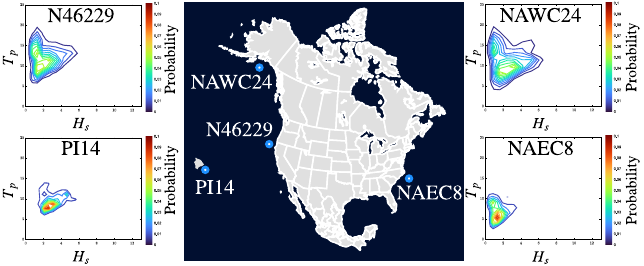}
        \caption{Joint distribution of $H_s$ and $T_p$ for the selected WEC farm site locations on Alaskan Coast, Pacific Coast, West Coast, and East Coast, based on data from~\cite{storlazzi2015future}.}
    \label{fig:locations}
    \vspace{-12pt}
\end{figure}

We describe the sea state using the JOint North Sea WAve Project (JONSWAP) spectrum, which is mathematically represented as a function of significant wave height $H_{s}$ and peak period $T_{p}$.
The incident wave field at the candidate sites is modeled through the superposition of a finite number of regular waves.
For each site of interest $L$, historical data for $H_{s}$ and $T_{p}$, collected from 1976 to 2005 \cite{storlazzi2015future} have been used within a Gauss quadrature approach.
Specifically, the probability matrix of the waves at each desired location and year $\mathbf{p}_{r}^{L}(H_{s},T_{p}| y)$ was estimated by inputting the associated Legendre-Gauss nodes corresponding to $H_{s}$ and $T_{p}$ in a kernel distribution function in \texttt{MATLAB}.
The location of these sites, along with their joint PDF for the first year of the study, is shown in Fig.~\ref{fig:locations}. 
Here, we assume that all candidate sites share the same water depth of $50~\unit{m}$, while maintaining their unique joint probability distribution for waves.

 \ifthenelse{\value{clearpageflag}>0}{\clearpage}{}
\section{Surrogate Model Validation}\label{sec:section3}

\begin{figure}[t]
    \captionsetup[subfigure]{justification=centering}
    \centering
    \begin{subfigure}{1\columnwidth}
    \centering
    \includegraphics[width=0.6\columnwidth]{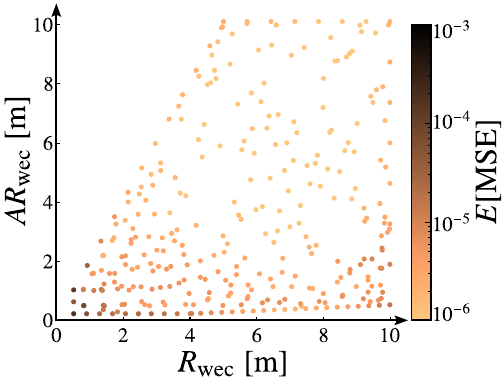}
    \caption{$B$.}
    \label{subfig:validation_B}
    \end{subfigure}
    \begin{subfigure}{1\columnwidth}
    \centering
    \includegraphics[width=0.6\columnwidth]{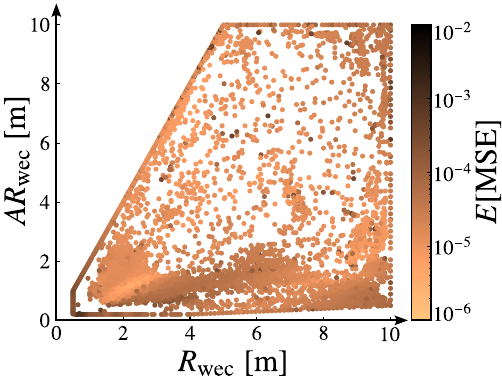}
    \caption{$A_{11}$.}
    \label{subfig:validation_A11}
    \end{subfigure}%
    \captionsetup[figure]{justification=centering}
        \caption{Mean squared error for the hydrodynamic coefficients 
        $B$ and $A_{11}$ averaged over all frequencies. (a) $B$ (selected from the single-WEC models), and (b) $A_{11}$ (selected from the two-WEC models).}
    \label{fig:validation}
    \vspace{-12pt}
\end{figure}

The surrogate model is a key piece of the proposed approach, and its accuracy will affect the predictions on the hydrodynamic coefficients and, subsequently, the generated power and the design optimization. 
Therefore, it is important to validate the performance of the surrogate model first.
The approach proposed in this study is subject to two sources of error: one that is inherent in developing ANNs and the other associated with the truncation order of MBE. 
In this section, we characterize the first source of error, which results in the presence of epistemic uncertainties \cite{azad2023overview}, associated with the lack of, or limitations on the availability of, training data.

In addition to metrics that are commonly used with ANNs (such as the mean-squared error (MSE) and the correlation coefficient), it is necessary to ensure that the models perform well across the entire design domain, particularly on or around the boundaries of the design space, to avoid extrapolation.
To this end, a MSE metric, averaged over all of the frequency-dependent outputs, was utilized to ensure reasonable model accuracy across the entire design domain.

The results of this analysis for two of the hydrodynamic coefficients ($B$ and $A_{11}$) are shown in Fig.~\ref{fig:validation}.
From this figure, it is clear that the average error is within the order of $10^{-2}$ and $10^{-3}$ for $B$ and $A_{11}$, respectively.
While not shown here, the same analysis was carried out for all of the developed models, and for the worst-performing models, an error bound of $10^{-2}$ was established. 
This analysis indicates that the surrogate models exhibit reasonable accuracy for most of the QoIs across the entire design domain.
Note that in Fig.~\ref{fig:validation}, the difference in the size of data is directly related to the use of QBC algorithm, in which an appropriate number of samples is intelligently selected based on the complexity of each model. 
The difference highlights the challenges in building an accurate surrogate model for $A_{11}$.

While the analysis presented above captures the scope of error in terms of hydrodynamic coefficients, there is still some ambiguity regarding the impact of these errors on power calculations.   
In other words, it is not clear how different combinations of error from surrogate modeling of hydrodynamic coefficients affect the generated power.
This is particularly important because it has the potential to negatively impact the optimized solution.  
In an initial effort to answer this question, we randomly selected $20,000$ samples from the entire design space, including plant ($\Rwec$, $\ARwec$), PTO control ($\BPTO$, $\KPTO$), and layout ($\AL$) for a $5$-WEC farm.
To isolate and separate the two sources of error discussed earlier, we consistently used MBE, but compared surrogate models with direct usage of multiple scattering.

\begin{figure}[t]
    \captionsetup[subfigure]{justification=centering}
    \centering
    \begin{subfigure}{\columnwidth}
    \centering
    \includegraphics[width=0.6\columnwidth]{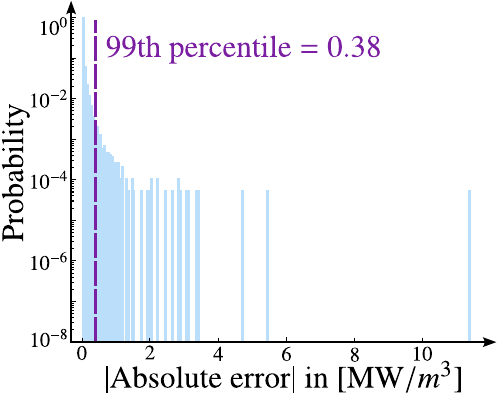}
    \caption{Error histogram.}
    \label{subfig:Error_log}
    \end{subfigure}
    \begin{subfigure}{\columnwidth}
    \centering
    \includegraphics[width=0.6\columnwidth]{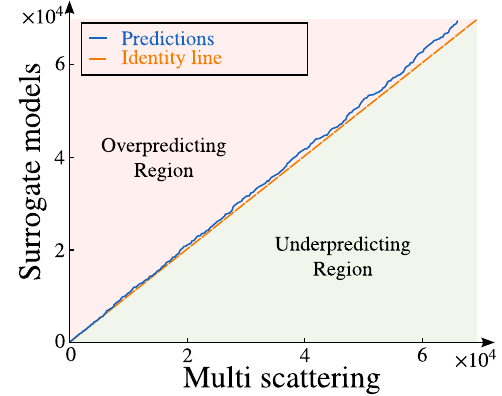}
    \caption{Energy estimate (\unit{MWh}).}
    \label{subfig:one2one}
    \end{subfigure}%
    \captionsetup[figure]{justification=centering}
    \caption{Characterization of error from surrogate modeling for $20,000$ randomly selected specifications for a 5-WEC farm.}
    \label{fig:Error}
    \vspace{-12pt}
\end{figure}

As shown in Fig.~\ref{subfig:Error_log}, results from this analysis indicate that $99$th percentile of error in power calculation corresponds to $0.38~[\unit{MW/m^3}]$.  
Figure \ref{subfig:one2one} highlights a direct comparison between surrogate models and MS.
Overall, the performance of the surrogate models are acceptable, although the models have a light tendency to over-predict at high powers. 
With the knowledge that the error is acceptably small, we carry out our investigations on concurrent plant, control, and layout optimization with site selection. 
Note that further investigation of the error stemming from surrogate models and the error associated with MBE truncation order is needed and will be explored in future work.%

\ifthenelse{\value{clearpageflag}>0}{\clearpage}{}
\section{Results}\label{sec:section4}

In this section, we start by formulating the concurrent control co-design with layout optimization and site selection problem and present some results for three case studies.
The open-source toolbox developed to run these studies is made publicly available in \cite{code}. 
Due to the complexity of the problem, a global search optimization algorithm is required. 
Genetic algorithm has proven to be promising in the design of WEC farms \cite{lyu2019optimization}, and thus, is used in this study.
The following studies progressively increase the level of integration, and, therefore, the complexity of the proposed problem.

The problem is formulated such that plant design variables $\bm{p} = [\Rwec,\ARwec]^{T}$ remain uniform, resulting in a farm with homogeneous devices.
However, PTO parameters $\bm{u} = [\BPTO,\KPTO]^{T}$ are first tuned for the entire farm in Case Study \rom{2}, and then optimized for each device in Case Study \rom{3}.
The location of the first device is fixed at the origin of the coordinate system and the farm area is restricted to the right half-plane through $0.5\sqrt{20000\Nwec}$.
A minimum distance $s_{d} =2\Rwec + 10~[\unit{m}]$ between the center of WEC devices is prescribed for maintenance ships to safely pass. 
The concurrent plant, control and layout optimization with site selection can now be formulated as:
\begin{subequations}
 \label{Eqn:OPtimization}
 \begin{align}
 \underset{\bm{p},\bm{u}, \AL}{\textrm{minimize:}}
 \quad & - p_{v}(\bm{p}, \bm{u}, \AL, L)   \label{Eqn:Obj} \\
 \textrm{subject to:} \quad
   \begin{split}
   &  2\Rwec + s_{d} - \bm{l}_{pq} \leq 0 \quad\\
        & \qquad \forall ~~  p,q = 1, 2, \dots, \Nwec \quad p \neq q \label{Eqn:distanceconst}
   \end{split} \\
&  \underaccent{\bar}{\bm{p}} \leq \bm{p} \leq \bar{\bm{p}}\label{Eqn:plantconst}\\
&  \underaccent{\bar}{\bm{u}} \leq \bm{u} \leq \bar{\bm{u}} \label{Eqn:controlconst}\\
&  \underaccent{\bar}{\AL} \leq \AL \leq \bar{\AL}\label{Eqn:layoutconst}\\
\textrm{where:} \quad & \bm{p} = [\Rwec, \ARwec]^{T} \in \mathbb{R}^{2} \notag\\
                & \bm{u} = [\KPTO, \BPTO]^{T} \in \mathbb{R}^{{2}\Nwec}  \notag \\
                & \AL = [\bm{x},\bm{y}] \in \mathbb{R}^{{2}(\Nwec-1)} \notag 
 \end{align} 
\end{subequations}
\noindent
where $L$ is the candidate site, affecting the optimization problem through the associated probability matrix $\mathbf{p}_{r}^{L}(H_{s},T_{p}, y)$, over $n_{yr} = 30$ years of the life of the farm. $\bar{\parm}$ and $\underaccent{\bar}{\parm}$ are lower and upper bounds on optimization variables. Table~\ref{Tab:Parameter_Opt} presents some key parameters used in this study.

 \begin{table}[t]
    \caption{Problem parameters.}
    \label{Tab:Parameter_Opt}
    \renewcommand{\arraystretch}{1.1}
    \setlength{\tabcolsep}{4pt}
    \centering
    \begin{tabular}{l c l c}
    \hline  \hline
    \textrm{\textbf{Parameter}} & \textrm{\textbf{Value}} &  \textrm{\textbf{Parameter}} & \textrm{\textbf{Value}} \\
    \hline
    $\Rwecmin$ & $0.5~[\unit{m}]$ & $\Rwecmax$ & $10~[\unit{m}]$ \\
    $\ARwecmin$ & $0.2~[\unit{m}]$ & $\ARwecmax$ & $10~[\unit{m}]$ \\
    $\Dwecmin$ & $0.5~[\unit{m}]$ & $\Dwecmax$ & $20~[\unit{m}]$ \\
    $\KPTOmin$ & $-5\times 10^{5}~[\unit{N/m}]$ & $\KPTOmax$ & $5\times 10^{5}~[\unit{N/m}]$ \\
    $\BPTOmin$ & $0~ [\unit{Ns/m}]$ & $\BPTOmax$ & $5\times 10^{5}~ [\unit{Ns/m}]$\\
    $\underaccent{\bar}{\bm{x}}$ & $0~[\unit{m}]$ &  $\bar{\bm{x}}$ & $0.5\sqrt{20000\Nwec}~[\unit{m}]$ \\
    $\underaccent{\bar}{\bm{y}}$ & $-0.5\sqrt{20000\Nwec}~[\unit{m}]$ &  $\bar{\bm{y}}$ & $0.5\sqrt{20000\Nwec}~[\unit{m}]$ \\
    $\rho$ & $1025~ [\unit{kg/m^3}]$ & $g$ & $9.81~[\unit{m/s^{2}}]$ \\
    $s_{d}$ & $10 ~[\unit{m}]$ & $\Nwec$ & $5$ \\
    $n_{yr}$   & $30~[\unit{years}]$ & $n_{r}$  & $200$\\
    $n_{gq}$   & $20$  & $\eta_{\text{pcc}}$ & $0.8$ \\
    $\eta_{\text{oa}}$ & $0.95$ & $\eta_{\text{t}}$ & $0.98$\\
    $\underaccent{\bar}{\omega}$ & $0.3$ & $\bar{\omega}$ & $2$\\
    \hline \hline
    \end{tabular}
    \vspace{-12pt}
\end{table}

\subsection{Study \rom{1}:Plant and Layout Optimization} 

In this case study, we assume that the PTO control parameters are fixed, and thus, the design variables in Eq.~(\ref{Eqn:OPtimization}) are reduced to [$\bm{p},\AL$].  
In addition to gaining a better understanding of the problem by comparing the results with solutions obtained using multi scattering, our goal here is to also assess the performance of the surrogate models in terms of their capability to capture the hydrodynamic interaction effect. 
To this end, we use the q-factor, which is defined as the ratio of the power from the whole farm to the sum of power from individual devices operating in isolation.
The study was investigated for all four of site locations, using $\KPTO = -5\times 10^{3}~[\unit{N/m}]$ and $\BPTO = 5\times 10^{5}~[\unit{Ns/m}]$.
The results are presented in Table~\ref{Tab:Study1}.
\begin{table}[t]
\renewcommand{\arraystretch}{1.2}
  \begin{threeparttable}[b]
  \begin{tabularx}{\linewidth}{@{}Y@{}}
   \caption{Optimized solution for Study~\rom{1}, consisting of plant and layout optimization for different site locations.}
   \label{Tab:Study1}
   \resizebox{\columnwidth}{!}{%
   \begin{tabular}{r r r r r r r r}
     \hline  \hline
    \multirow{2}{*}{\rotatebox[origin=c]{0}{\textbf{Site}}} & \multirow{2}{*}{\textrm{\textbf{Method}}} & \multicolumn{2}{c}{\textrm{\textbf{Plant}}} &  \multirow{2}{*}{\textrm{\textbf{L}\tnote{b}}}& \multirow{2}{*}{\textrm{{P}\tnote{c}}} & \multirow{2}{*}{$q$\tnote{d}} & \multirow{2}{*}{\textrm{\textbf{Time}\tnote{e}}}\\ \cline{3-4}
     & & $\Rwec$\tnote{a} & $\ARwec$ & & & &  \\
     \hline 
     \multirow{2}{*}{\rotatebox[origin=c]{0}{\textrm{NAWC24}}} & $\text{SM-MBE}$ & $3.00$ & $6.00$ & \parbox[t]{2mm}{\multirow{5}{*}{\rotatebox[origin=c]{90}{\textrm{Fig.~\ref{fig:PLCaseStudyI}}}}}& $160.73$ & $0.99$  & $2.6$\\
    & $\text{MS}$ & $3.05$ & $6.1$ & & $168.87$ & $1.01$ & $87.9$ \\
      \cline{1-4}\cline{6-8}
    \multirow{1}{*}{\rotatebox[origin=c]{0}{\textrm{NAEC8}}} & $\text{SM-MBE}$ & $3.70$ & $7.41$ & & $16.17$ & $1.02$  & $3.4$\\
    \cline{1-4}\cline{6-8}
    \multirow{1}{*}{\rotatebox[origin=c]{0}{\textrm{PI14}}} & $\text{SM-MBE}$ & $3.33$ & $6.66$ & & $33.53$ & $1.00$  & $3.5$\\
      \cline{1-4}\cline{6-8}
    \multirow{1}{*}{\rotatebox[origin=c]{0}{\textrm{N46229}}} & $\text{SM-MBE}$ & $2.91$ & $5.83$ & & $70.35$ & $1.00$  & $3.5$\\
    \hline \hline
    \end{tabular}%
    }
    \end{tabularx}
     \begin{tablenotes} [para,flushleft]
       \item [a] Calculated in [\unit{m}] \item [b] Optimized layout \item [c] Power calculated over farm's lifetime using MS in [\unit{MW}] \item [d] q-factor calculated using MS \item [e] Computational time per generation in [\unit{h}] 
     \end{tablenotes}
  \end{threeparttable}
    \vspace{-12pt}
\end{table}

From Table~\ref{Tab:Study1}, it is clear that both of the plant optimization variables change with each site location. 
With the exception of the N46229 site location, the WEC radius $\Rwec$ increases with the decrease in the wave energy resource of the site, indicating a $12.35\%$ change between the highest and lowest radius.
While the slender ratio $\ARwec$ of the WEC changes with location, its draft remains very close to $0.5~[\unit{m}]$ in all cases, indicating that under the current assumptions (and objective function), a smaller draft is the most optimal for all locations.
Note that the optimal plant solution obtained using MS (only for NAWC24, due to high computational cost) closely matches that found by our surrogate models.

\begin{figure*}[t]
    \captionsetup[subfigure]{justification=centering}
    \centering
    \begin{subfigure}{\columnwidth}
    \centering
    \includegraphics[scale=0.7]{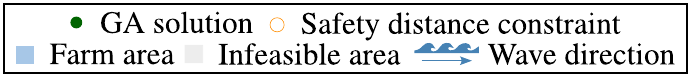}
    \label{subfig:Legend}
    \end{subfigure}
    \begin{subfigure}{0.5\columnwidth}
    \centering
    \includegraphics[scale=0.8]{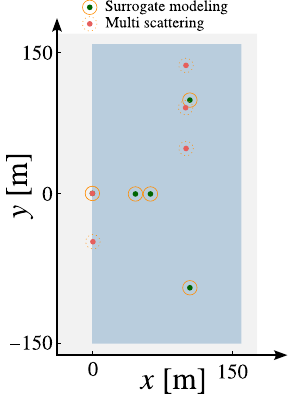}
    \caption{Alaska Coast (NAWC24).}
    \label{subfig:LPAlaska}
    \end{subfigure}%
    \begin{subfigure}{0.5\columnwidth}
    \centering
    \includegraphics[scale=0.8]{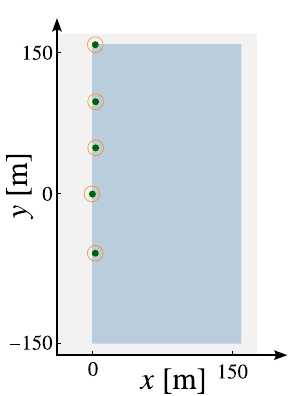}
    \caption{East Coast (NAEC8).}
    \label{subfig:LPEast}
    \end{subfigure}%
    \begin{subfigure}{0.5\columnwidth}
    \centering
    \includegraphics[scale=0.8]{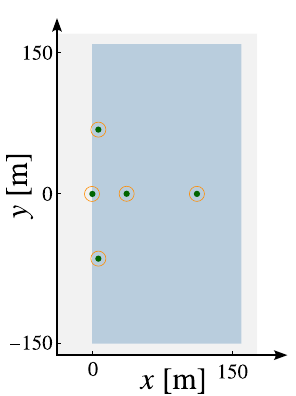}
    \caption{Pacific Islands (PI14).}
    \label{subfig:LPPacific}
    \end{subfigure}%
    \begin{subfigure}{0.5\columnwidth}
    \centering
    \includegraphics[scale=0.8]{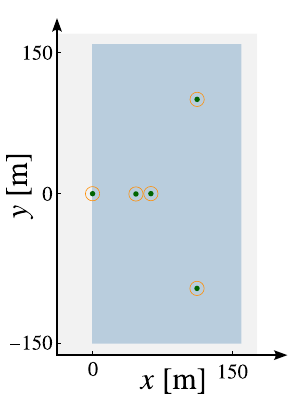}
    \caption{West Coast (N46229).}
    \label{subfig:LPWest}
    \end{subfigure}%
    \captionsetup[figure]{justification=centering}
    \caption{Optimized layout for Study \rom{1}: plant and layout optimization (drawn in scale).} 
    \label{fig:PLCaseStudyI}
    \vspace{-12pt}
\end{figure*}

The optimized location of the farm for each site location is shown in Fig.~\ref{fig:PLCaseStudyI}.
These figures highlight the impact of site location on the optimized layout of the farm.
Particularly, for NAWC24 and N46229 locations, which are classified as high wave energy resource regions, the optimized array is closely similar, with symmetry with respect to the direction of the incident wave.
The optimized layout for the site location with the lowest wave energy resource, i.e.,~NAEC8 is an array vertical to the direction of the wave travel.
The optimized layout for PI14, which is considered a low/medium resource, is also symmetrical with respect to the direction of wave travel, albeit different from the other cases.
Note that the optimized layout obtained using MS (shown in Fig.~\ref{subfig:LPAlaska}) is different from that obtained using surrogate models. 
In this case, the MS layout creates two clusters of WEC devices.

As expected, under the assumption of a fixed controller, the average power captured by the farm at each location is proportional to the available resource in that region. The q-factor remains extremely close to $1$ for all of the cases.
This can be attributed to the fact that irregular waves along with the consideration of their probabilities, result in a smoothing effect that lessens the q-factor values compared to the regular waves \cite{Han2023}.
The computational time of the investigation is also reported per generation of the genetic algorithm.
Using surrogate models with MBE to estimate the hydrodynamic coefficients has resulted in over $33$ times improvement in the computational time.

\subsection{Study \rom{2}: Plant, Farm-level Control, and Layout Optimization with Site Selection}
This section presents some results for the plant, control, and layout optimization with site selection, under the assumption that the PTO parameters of the WEC devices are uniform across the farm. 
Therefore, this case study assumes that $\bm{u} = [\KPTO, \BPTO]^{T} \in \mathbb{R}^{{2}}$.
Due to the high computational cost of directly using MS, this study is only investigated using the SM-MBE approach, using genetic algorithm. 
Note, however, that when control variables are included in the optimization problem, the optimal control is often a resonator.
This may result in the amplification of device motions, leading to unrealistic power outputs. 
One remedy to address this challenge is to use time-domain models with more practical control algorithms and device motion constraints.
This, however, is outside of the scope of the current study.
The optimized solutions for all candidate sites are presented in Table~\ref{Tab:Study2}. 

\begin{table}[t]
\renewcommand{\arraystretch}{1.1}
  \begin{threeparttable}[b]
  \begin{tabularx}{\linewidth}{@{}Y@{}}
   \caption{Optimized solution for Study~\rom{2}, consisting of plant, farm-level control, and layout optimization with site selection.}
   \label{Tab:Study2}
   \centering
   \resizebox{\columnwidth}{!}{%
   \begin{tabular}{r r r r r r r r r}
     \hline  \hline
    \multirow{2}{*}{\textrm{\textbf{Site}}} & \multicolumn{2}{c}{\textrm{\textbf{Plant}}} &  \multicolumn{2}{c}{\textrm{\textbf{Control}\tnote{b}}}& \multirow{2}{*}{\textrm{\textbf{L}\tnote{c}}} & \multirow{2}{*}{\textrm{{P}}\tnote{d}} & \multirow{2}{*}{\textrm{\textbf{Time}\tnote{e}}} \\ \cline{2-3} \cline{4-5}
     & $\Rwec\tnote{a}$ & $\ARwec$ & $\BPTO$ & $\KPTO$  & &  & \\
          \hline 
     $\text{NAWC24}$ & $3.60$ & $6.95$ & $8.46$ & $-367.00$ &  \parbox[t]{2mm}{\multirow{4}{*}{\rotatebox[origin=c]{90}{\textrm{Fig.~\ref{fig:OptimalArray_CaseII}}}}} & $7258$  & $2.6$ \\
     $\text{NAEC8}$ & $3.74$ & $6.68$ & $6.22$ & $-368.76$ &  & $1703$ & $2.3$ \\
     $\text{PI14}$ & $3.89$ & $7.44$ & $17.99$& $-379.71$ &  & $1166$ & $2.4$ \\
     $\text{N46229}$ & $3.65$ & $7.07$ & $8.11$& $-385.31$ &  & $3992$ & $2.9$ \\
    \hline \hline
    \end{tabular}%
}
\end{tabularx}
     \begin{tablenotes} [para,flushleft]
        \item [a] Calculated in [\unit{m}] \item [b] Calculated in [\unit{kNs/m}] and [\unit{kN/m}] \item [c] Optimized layout \item [d] Power calculated in [\unit{MW}] over the lifetime of the farm using MS \item [e] Computational time per generation in [\unit{h}] 
     \end{tablenotes}
  \end{threeparttable}
    \vspace{-12pt}
\end{table}

From these results, it is evident that the optimal plant variables change for different site locations.
Specifically, the smallest WEC radius corresponds to the Alaskan coast (NAWC24), followed by that of the West Coast.
This indicates that the WEC radius tends to be smaller for high wave-resource regions, and increases with decreasing levels of wave energy resource. 
Note that the WEC draft dimension remains close to $0.5~[\unit{m}]$, although it changes in the range of $0.51$ to $0.56$.
Further investigations are required to clarify if this observation is associated with the site location, the choice of the objective function, or a result of the optimization algorithm and/or hydrodynamic surrogate models.

\begin{figure*}[t]
\centering
\captionsetup[subfigure]{justification=centering}
\centering
\begin{subfigure}{0.25\textwidth}
\centering
\includegraphics[scale=0.8]{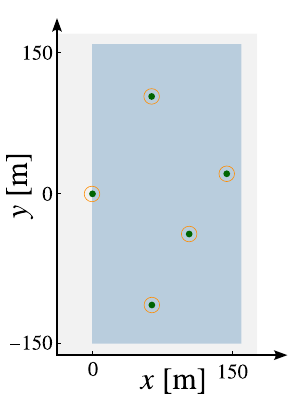}
\caption{Alaska Coast (N46229).}
\label{fig:AlaskaII}
\end{subfigure}%
\begin{subfigure}{0.25\textwidth}
\centering
\includegraphics[scale=0.8]{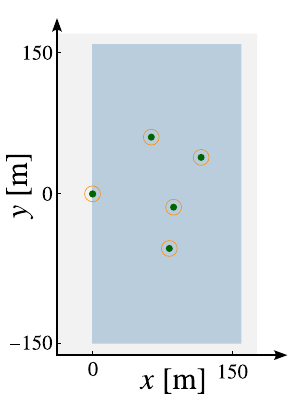}
\caption{East Coast (NAEC8).}
\label{fig:EastII}
\end{subfigure}%
\begin{subfigure}{0.25\textwidth}
\centering
\includegraphics[scale=0.8]{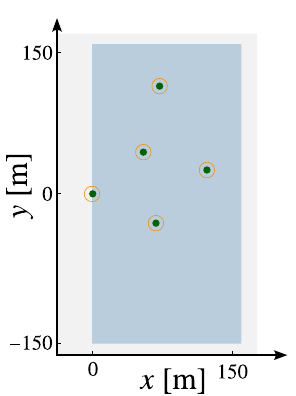}
\caption{Pacific Islands (PI14).}
\label{fig:PacificII}
\end{subfigure}%
\begin{subfigure}{0.25\textwidth}
\centering
\includegraphics[scale=0.8]{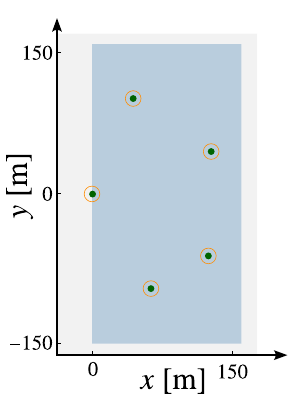}
\caption{West Coast (N46229).}
\label{fig:WestII}
\end{subfigure}%
\captionsetup[figure]{justification=centering}
\caption{Optimized layout for Study \rom{2}: plant, farm-level control, and layout optimization (drawn in scale).} 
\label{fig:OptimalArray_CaseII}
\vspace{-12pt}
\end{figure*}

The optimized PTO parameters also change for various site locations.
These parameters exhibit a maximum change of $65.43\%$ and $4.75\%$ percent for PTO damping and stiffness, respectively.
While the results do not indicate a uniform trend with respect to the region's energy resources, they emphasize the need to design the WEC for specific site locations because the design of control parameters is largely affected by dominant wave frequencies in the specific region.
The optimized layout of the farm is shown in Fig.~\ref{fig:OptimalArray_CaseII}, indicating different layout configurations for the farm.
Note that when control optimization is considered, the optimized layouts start to lose symmetry.
This can be associated with the fact that an optimal controller is often a resonator for a single device in the farm, resulting in large amplitudes of the device motion that deviate from the otherwise existing symmetry under the assumption of a fixed, non-resonant controller.

Note that we have excluded the q-\text{factor} calculations from Table~\ref{Tab:Study2}.
This is because q-factor has certain limitations in assessing the performance of the farm when control optimization is involved.
In other words, due to the resonance characteristics of an optimal controller, and the the lack of any realistic power limits on the PTO system in this study, the numerator of the q-\text{factor} will be large.
However, in the denominator, the controller will be off-resonance (due to the device's natural frequency being dependent on the added mass matrix).
Therefore, assessing the performance of the farm using the traditional q-\text{factor} is misleading when control optimization is involved.

\begin{table}[t]
\renewcommand{\arraystretch}{1.1}
  \begin{threeparttable}[b]
  \begin{tabularx}{\linewidth}{@{}Y@{}}
   \caption{Optimized solution for Study \rom{3}, consisting of plant, device-level control, and layout optimization with site selection.}
   \label{Tab:Study3}
   \centering
      \resizebox{\columnwidth}{!}{%
   \begin{tabular}{r r r r r r r}
     \hline  \hline
    \multirow{2}{*}{\textrm{\textbf{Site}}} & \multicolumn{2}{c}{\textrm{\textbf{Plant}}} &  \multirow{2}{*}{\textrm{\textbf{Control}}} & \multirow{2}{*}{\textrm{\textbf{L}\tnote{a}}}  & \multirow{2}{*}{\textrm{{P}}\tnote{b}} & \multirow{2}{*}{\textrm{\textbf{Time}\tnote{c}}} \\ \cline{2-3}
     & $\Rwec~[\unit{m}]$ & $\ARwec$ & & & & \\
     \hline 
     $\textrm{NAWC24}$ & $3.88$ & $7.70$ & \parbox[t]{2mm}{\multirow{4}{*}{\rotatebox[origin=c]{90}{\textrm{Fig.~\ref{fig:FarmControl}}}}} & \parbox[t]{2mm}{\multirow{4}{*}{\rotatebox[origin=c]{90}{\textrm{Fig.~\ref{fig:OptimalArray_CaseIII}}}}} & $4131$  &  $4.6$\\
     $\text{NAEC8}$ & $3.82$ & $7.11$ & &  & $59$ & $2.3$ \\
     $\text{PI14}$ & $3.86$ & $7.21$ & & & $428$ & $2.1$ \\
     $\text{N46229}$ & $2.85$ & $5.65$ & &  & $1197$ & $2.8$ \\
    \hline \hline
    \end{tabular}%
    }
    \end{tabularx}
     \begin{tablenotes} [para,flushleft]
       \item [a] Optimized layout \item [b] Power calculated  over farm's lifetime in [\unit{MW}] \item [c] Computational time per generation in [\unit{h}]
     \end{tablenotes}
  \end{threeparttable}
    \vspace{-12pt}
\end{table}

\subsection{Study \rom{3}: Plant, Device-level Control, and Layout Optimization with Site Selection}
In this case study, we investigate the impact of selecting optimized control parameters for each individual WEC device on the overall performance of the farm.
Compared to Study II, where all WEC devices shared the same $\BPTO$ and $\KPTO$, this study allows these control parameters to vary across the farm; thus, $\bm{u} = [\KPTO, \BPTO]^{T} \in \mathbb{R}^{{2\Nwec}}$.
Using GA, the results for this case study are shown in Table~\ref{Tab:Study3}.

While further investigations are required to highlight practical considerations for the control strategy, this study allows us to assess the impact of device-level optimized control on power generation. 
Theoretically, the device-level control co-design and layout optimization of the farm has the potential to result in increased power generation and improved performance of the farm. 
Note that compared to Study \rom{2}, which has a total of $12$ optimization variables, the size of the problem grows to $20$ for Study \rom{3}.
Therefore, it is expected to witness some increase in the computational time.
This is, however, true only for NAWC24, as all the other cases converged relatively faster.

\begin{figure}
    \centering
    \includegraphics[width=0.6\columnwidth]{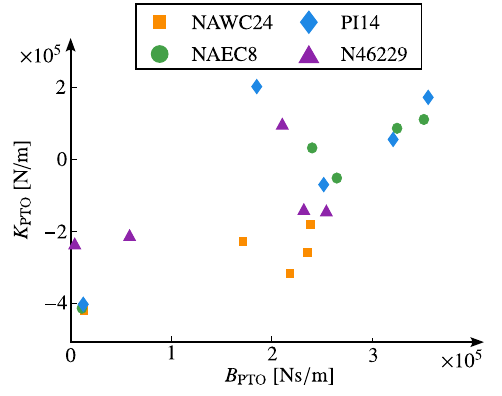}
    \caption{Optimized PTO control parameters for the WEC farm at candidate sites.}
    \label{fig:FarmControl}
    \vspace{-12pt}
\end{figure}

\begin{figure*}[t]
\centering
\captionsetup[subfigure]{justification=centering}
\centering
\begin{subfigure}{0.25\textwidth}
\centering
\includegraphics[scale=0.8]{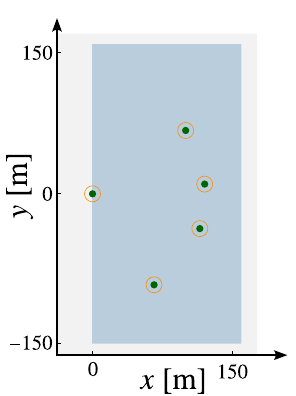}
\caption{Alaska Coast (N46229).}
\label{fig:AlaskaIII}
\end{subfigure}%
\begin{subfigure}{0.25\textwidth}
\centering
\includegraphics[scale=0.8]{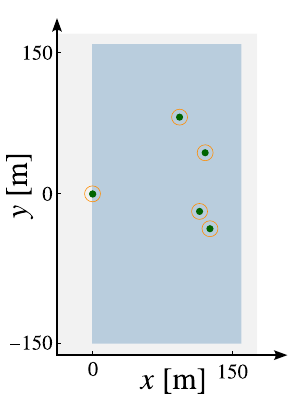}
\caption{East Coast (NAEC8).}
\label{fig:EastIII}
\end{subfigure}%
\begin{subfigure}{0.25\textwidth}
\centering
\includegraphics[scale=0.8]{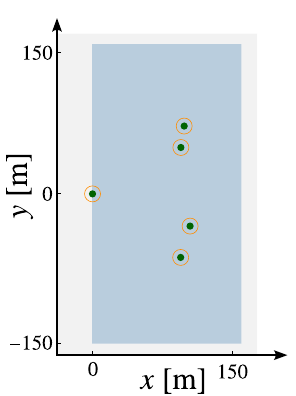}
\caption{Pacific Islands (PI14).}
\label{fig:PacificIII}
\end{subfigure}%
\begin{subfigure}{0.25\textwidth}
\centering
\includegraphics[scale=0.8]{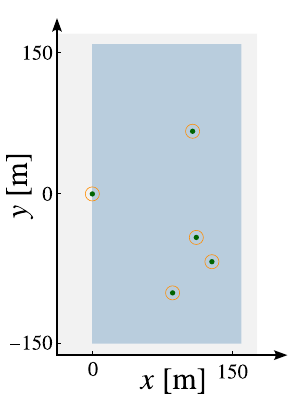}
\caption{West Coast (N46229).}
\label{fig:WestIII}
\end{subfigure}%
\captionsetup[figure]{justification=centering}
\caption{Optimized layout for Study \rom{3}: plant, device-level control, and layout optimization (drawn in scale).} 
\label{fig:OptimalArray_CaseIII}
\vspace{-12pt}
\end{figure*}

The results indicate a $26.55\%$ change between the smallest and largest WEC radius, corresponding to N46229 and NQWC24, respectively.
Similar to the previous cases, the draft dimension of the WEC devices remains close to $0.5~[\unit{m}]$.
The control parameters are scattered within the control design space, as shown in Fig.~\ref{fig:FarmControl}.
The optimized layouts for this case study are shown in Fig.~\ref{fig:OptimalArray_CaseIII}.
Once again, while some symmetry is observed, there is no regular layout witnessed in the results.
It is also noticeable that, for the two low- and medium-energy resource regions, the WEC devices seem to be clustered strategically in pairs (Figs.~\ref{fig:EastIII} and \ref{fig:PacificIII}).
Note that the power generated in this study is smaller than that of Case Study \rom{2}.
This outcome seems to be an artifact of the optimizer or the potential amplification of error from surrogate models.
One remedy to address this challenge is to use GA's solution as a starting point to solve the optimization problem with a gradient-based optimizer. 
This hybrid approach will be pursued in the future of this research.

\begin{figure*}[t]
\centering
\captionsetup[subfigure]{justification=centering}
\centering
\begin{subfigure}{0.33\textwidth}
\centering
\includegraphics[scale=0.7]{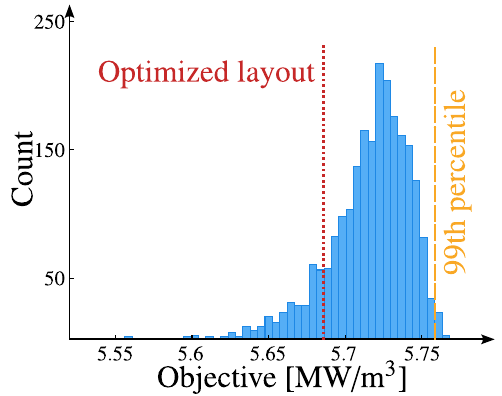}
\caption{Study \rom{1}: Concurrent plant and layout optimization.}
\label{fig:Analysis_1_Alaska}
\end{subfigure}%
\begin{subfigure}{0.33\textwidth}
\centering
\includegraphics[scale=0.7]{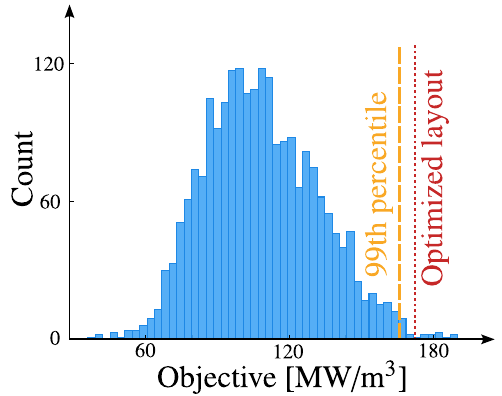}
\caption{Study \rom{2}: Concurrent plant, farm-level control and layout optimization.}
\label{fig:Analysis_2_Alaska}
\end{subfigure}%
\begin{subfigure}{0.33\textwidth}
\centering
\includegraphics[scale=0.7]{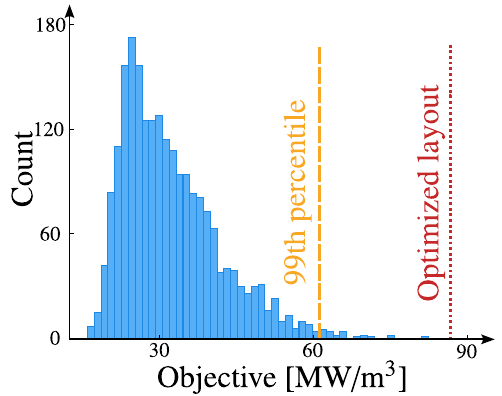}
\caption{Study \rom{3}: Concurrent plant, device-level control and layout optimization.}
\label{fig:Analysis_3_Alaska}
\end{subfigure}%
\captionsetup[figure]{justification=centering}
\caption{Comparison between generated power using select optimized layout solutions from Studies \rom{1}, \rom{2}, and \rom{3} to $2500$ randomly-selected layouts simulated using multi scattering all at the Alaskan Coast (NAWC24).} 
\label{fig:Analysis_app}
\vspace{-12pt}
\end{figure*}

\ifthenelse{\value{clearpageflag}>0}{\clearpage}{}
\section{Conclusion }\label{sec:conclusion}
To better capitalize on domain coupling in the design of WEC devices, we formulated and implemented a concurrent plant, control, and layout optimization with considerations for variations in environmental inputs resulting from different site locations.
To efficiently estimate the hydrodynamic coefficients, we used surrogate models constructed by training ANN on through an active learning strategy that is known as query by committee.
The surrogate models were validated across the input space, and the error was characterized in terms of the energy generated over the farm's lifetime.
The results indicated strong coupling between plant, control, layout, and environment, signifying the importance of domain coupling in the design of WEC farms.

While the study shows promising directions towards the system-level optimization of WEC farms, more effort is required to first, improve the control strategy to prevent unrealistic device motions.
This approach is expected to address some of the concerns regarding the amplification of errors from surrogate models when control optimization is involved.
Addressing these limitations will be the focus of our future work.

\vspace{-3pt}
\begin{acknowledgment}
\vspace{-3pt}
The authors gratefully acknowledge the financial support from National Science Foundation Engineering Design and Systems Engineering Program, USA under grant number CMMI-2034040.
\end{acknowledgment}

 \nocite{*}
\bibliographystyle{IEEEtran}
\bibliography{bib}

\section{Appendix}\label{sec:Appendix}

This appendix includes further analysis of some of the studies presented in the main body of the manuscript.
Due to the existing page limitations, this section was not included in the American Control Conference submission. 
Our goal here is to compare the optimized layout solutions, obtained using surrogate modeling in combination with MBE with randomly-selected layouts, simulated using multi scattering.
To this end, we first prescribed the optimized plant and control variables obtained from each case study for the simulation.
Next, $2500$ randomly-selected layouts that satisfy the problem constraints were generated and simulated using multi scattering approach.
The objective function estimates from these simulations are then used along with that from the optimized solution to generate histogram plots.   
The results from this analysis are presented in Fig.~\ref{fig:Analysis_app}.

\begin{figure}
\captionsetup[subfigure]{justification=centering}
\centering
\begin{subfigure}{\columnwidth}
\centering
\includegraphics[width = 0.8\columnwidth]{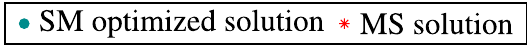}
\label{subfig:L_opt_Analysis2_leg}
\end{subfigure}
\begin{subfigure}{0.5\columnwidth}
\centering
\includegraphics[width = \columnwidth]{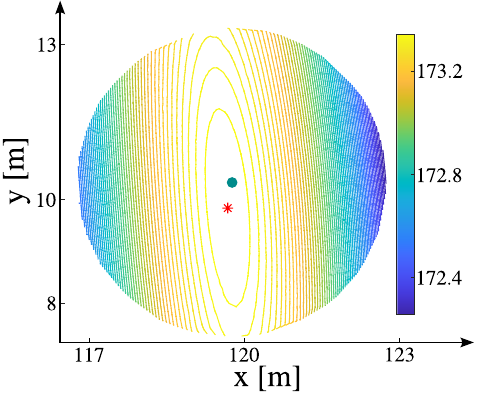}
\caption{WEC \# 2}
\label{subfig:L_opt_Analysis2_WEC2}
\end{subfigure}%
\begin{subfigure}{0.5\columnwidth}
\centering
\includegraphics[width = \columnwidth]{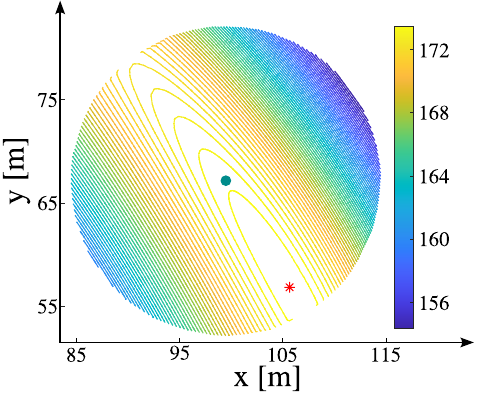}
\caption{WEC \# 3}
\label{subfig:L_opt_Analysis2_WEC3}
\end{subfigure}
\captionsetup[figure]{justification=centering}
\centering
\caption{Sensitivity Analysis of the optimized layout from Case Study \rom{3} at the Alaska Coast for WEC \# 2 and WEC \# 3. The objective function contours are created using a MS approach, with a large number of random simulations in the neighborhood of the WEC devices. The MS solution is obtained by keeping all other WECs at their optimized locations.}
\label{fig:L_opt_Analysis2}
\end{figure}

From Fig.~\ref{fig:Analysis_1_Alaska}, it is clear that the optimized solution obtained using surrogate modeling with MBE in Study \rom{1} can be improved, as it is lower than the $99$th percentile of the simulated data.
Note, however, that the magnitude of this difference is relatively small, $1.27\%$ in the objective function value, or approximately $2~[\unit{MW}]$ over $30$ years of the farm life.
Evidenced by Figs.~\ref{fig:Analysis_2_Alaska} and \ref{fig:Analysis_3_Alaska}, this is not the case when control optimization is included in the investigation (i.e.,~Study \rom{2}, and \rom{3}).
In fact, for both of these studies, the optimized solution is higher than $99$th percentile of the simulated power, indicating that the optimized layouts are indeed effective in improving farm performance.
This is directly related to the fact that the optimal controller for the heaving cylinder WEC device is often a resonator, and the natural frequency of the devices within the farm are affected by the layout. 
Therefore, when the layouts are randomly selected, the controller is no longer optimal for the farm.
This observation, which results in lower power generation from the simulated layouts, highlights the prominent role of control optimization at the farm level for WEC devices. 

While these investigations answer some questions regarding the effectiveness of the proposed surrogate modeling strategy, it is still unclear whether or not the surrogate models are capable of finding an optimal layout solution.
To answer this question, we perform a sensitivity analysis of the optimized layout solution obtained using surrogate modeling.
Specifically, using the optimal plant and control values from Case Study \rom{3} in the Alaska Coast as prescribed parameters, we perturb the position of one of the WECs within the farm, while keeping the remaining WECs at their optimized locations.
Simulating a large number of perturbations in the neighborhood of the WEC of interest using MS, we can assess the proximity of the optimized solution to the true MS solution.  
The sensitivity analysis results, along with the MS solution, for WEC \# 2 and \# 3 are presented in Fig.~\ref{fig:L_opt_Analysis2}.

From Fig.~\ref{subfig:L_opt_Analysis2_WEC2}, it is clear that the optimized solution obtained through the surrogate modeling strategy is very close to the MS solution.
For WEC \# 3, however, the optimized solution is approximately $12~[\unit{m}]$ away from the MS solution. 
Note, however that despite this distance, the SM objective only differs by $1.18\%$ from the MS solution. 
Overall, these results indicate that while the performance of the surrogate models is acceptable, further improvements can be achieved by using a hybrid optimization strategy, in which GA's solution, obtained using surrogate modeling with MBE, is a starting point to solve an optimization problem using a gradient-based optimizer with multi scattering.
This approach, which will be investigated in our future work, offers a promising avenue to reduce the impact of errors inherent in the proposed methodology, resulting from epistemic uncertainties in surrogate modeling, as well as the truncation order of MBE.

\end{document}